\documentclass[3p,times, twocolumn, authoryear]{elsarticle}
\graphicspath{ {./figures/} }
\usepackage{hyperref}
\usepackage{float}
\usepackage{verbatim} 
\usepackage{apalike}
\restylefloat{figure}
\restylefloat{table}
\usepackage{graphicx}
\usepackage{multirow}
\usepackage{amsmath}
\usepackage{booktabs}
\usepackage{algorithm}
\usepackage{algorithmic}
\usepackage{tabularx}
\usepackage{xcolor}
\usepackage{amssymb}
\usepackage[figuresright]{rotating}
\usepackage{makecell}
\usepackage{lipsum}

\newcommand{\yj}[1]{\textcolor{black}{#1}}

\journal{Expert Systems with Applications}

\biboptions{authoryear}

\begin{document}
\begin{frontmatter}

\begin{titlepage}
\begin{center}
\vspace*{1cm}

\textbf{ \large An Adaptive Dual-level Reinforcement Learning Approach for Optimal Trade Execution}

\vspace{1.5cm}

Soohan Kim$^b$ (zeanvszed@g.skku.edu), Jimyeong Kim$^b$(jimkim@g.skku.edu), Hong Kee Sul$^{a}$ (hksul@cau.ac.kr), Youngjoon Hong$^b$ (hongyj@g.skku.edu) \\

\hspace{10pt}

\begin{flushleft}
\small  
$^a$ Department of Finance, Chung-Ang University, Seoul, Republic of Korea. \\
$^b$ Department of Mathematics, Sungkyunkwan University, Suwon, Republic of Korea

\vspace{1cm}
\textbf{Corresponding Author:} \\
Youngjoon Hong \\
Department of Mathematics, Sungkyunkwan University, Suwon, Republic of Korea\\
Email: hongyj@g.skku.edu\\
\vspace{2mm}
Hong Kee Sul \\
Department of Finance, Chung-Ang University, Seoul, Republic of Korea.\\
Email: hongyj@g.skku.edu

\end{flushleft}        
\end{center}
\end{titlepage}

\title{An Adaptive Dual-level Reinforcement Learning Approach for Optimal Trade Execution}

\author[label2]{Soohan Kim \corref{cor1}}
\ead{zeanvszed@g.skku.edu}

\author[label2]{Jimyeong Kim \corref{cor1}}
\ead{jimkim@g.skku.edu}

\author[label1]{Hong Kee Sul \corref{cor2}}
\ead{hksul@cau.ac.kr}

\author[label2]{Youngjoon Hong \corref{cor2}}
\ead{hongyj@g.skku.edu}

\cortext[cor1]{these authors contributed equally to this work.}
\cortext[cor2]{these authors contributed equally to this work, corresponding author}
\address[label1]{Department of Finance, Chung-Ang University, Seoul, Republic of Korea}
\address[label2]{Department of Mathematics, Sungkyunkwan University, Suwon, Republic of Korea}

\begin{abstract}
The purpose of this research is to devise a tactic that can closely track the daily cumulative volume-weighted average price (VWAP) using reinforcement learning.
Previous studies often choose a relatively short trading horizon to implement their models, making it difficult to accurately track the daily cumulative VWAP since the variations of financial data are often insignificant within the short trading horizon.
In this paper, we aim to develop a strategy that can accurately track the daily cumulative VWAP while minimizing the deviation from the VWAP.
We propose a method that leverages the U-shaped pattern of intraday stock trade volumes and use Proximal Policy Optimization (PPO) as the learning algorithm.
Our method follows a dual-level approach: a Transformer model that captures the overall(global) distribution of daily volumes in a U-shape, and a LSTM model that handles the distribution of orders within smaller(local) time intervals.
The results from our experiments suggest that this dual-level architecture improves the accuracy of approximating the cumulative VWAP, when compared to previous reinforcement learning-based models.
\end{abstract}


\end{frontmatter}

\section{Introduction}\label{introduction}
The optimal trade execution problem aims to find a strategy to optimally trade large orders within a given period of time. One of the most common and practical methods that practitioners frequently use is known as Volume Weighted Average Price (VWAP) trading. VWAP is calculated by adding up the dollars traded for every transaction (price multiplied by the number of shares traded) and then dividing by the total shares traded for the day. This gives an average price that takes into account both the price and the volume of shares traded. Funds and traders often use VWAP as a benchmark to compare the price at which they executed trades to the overall market price for the security, in order to evaluate the performance of their trades.

Recently, the use of Reinforcement Learning (RL) in optimal trade execution has become popular as it is a type of stochastic decision-making process that automates the practitioner's task of using past data to make decisions on when to execute orders.
Since RL approaches have many advantages, e.g. capable of capturing the market's microstructure, the trading results conducted by RL solutions often outperform traditional VWAP tracking approaches such as \cite{Bialkowski08} and \cite{Podobnik09}.

\begin{figure}[h]
\centering
 \includegraphics[width=7.5cm]{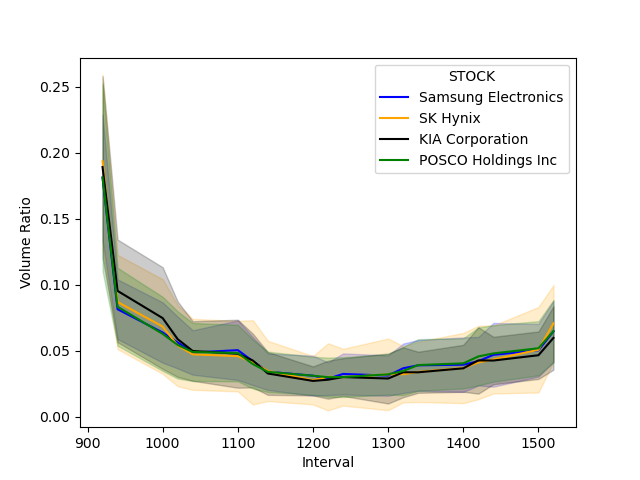}
\caption{Trade volume ratio over each 20-minute period throughout the day. The lines represent the 1-year averages of the ratios and the shaded regions are drawn from daily deviations from the averages.}\label{ushape}
\end{figure}

This paper aims to develop a strategy that can consistently track the daily cumulative VWAP. To achieve this goal, we start with the observation of an important stock trading characteristic, the U-shaped intraday trading pattern, documented in  \cite{JainJoh88} and \cite{goohartOhara97}. This is a well-known stylized fact stating that a stock's trading volume often follows a U-shaped pattern throughout the day. This means that volume is highest at the opening, 
falls rapidly to lower levels, and then rises again towards the close of the market, but is relatively low in the afternoon.
(See Figure \ref{ushape}).
In view of this observation, we propose a novel dual-level approach to minimize the market impact and track the daily cumulative VWAP accurately and consistently. As the name suggests, our model consists of two stages. In the first stage, the model decides how much trade volume will be executed in each interval, using the U-shape property as a guideline. In the second stage, the RL model decides how the orders will be executed in each interval. We propose two methods for the first stage, the statistical U-shape method and the U-shape Transformer method. The statistical U-shape method allocates the volume based on the historical average trade volumes. However, this approach cannot fully capture the day-to-day variations (See Figure \ref{ushape}). Alternatively, we  propose the U-shape Transformer for figuring out these variations. In the next stage, we apply LSTMs and PPO to distribute orders efficiently within each interval.

We expect that this dual-level approach improves the accuracy of approximating the cumulative VWAP by properly distributing the total order. Our contributions are as follows:
\begin{enumerate}
    \item We propose a novel approach for optimal trade execution that utilizes a dual-level strategy. Orders are allocated through two stages:  in the first stage we utilize the U-shaped pattern of intraday volumes and in the second stage we implement deep reinforcement learning.
    \item We further propose the U-shape Transformer model to capture the day-to-day oscillations of the intraday U-shape distribution.
 \item We conduct ablation studies to show the strengths of our methods in approximating the daily cumulative VWAP and reveal that utilizing deep reinforcement learning for short trading horizons is ineffective.
\end{enumerate}

\section{Backgrounds}
In this section, we discuss the basics of the limit order book (LOB) and the optimal trade execution process. 
Most modern financial exchanges, including the Korea Stock Exchange(KRX), offer electronic trading platforms with access to limit order books.
Traders often rely on the information provided by the limit order book to develop successful trading strategies.

\subsection{Limit Order Book}
In financial markets, a limit order is an order to buy or sell a fixed number of shares at a specified price. This order becomes part of the limit order book, which  records all limit orders for a specific stock. The bid price is the highest price a buyer is willing to pay for the security, and the ask price is the lowest price in which the seller is willing to sell. A market order, by contrast, is an order to buy or sell a fixed number of shares at the current market price without stating the price. When a market order is placed, it is matched by the best available limit orders in the LOB. The LOB is important for optimal trade execution and helps traders track supply and demand for a stock, making it easier to identify the best time and price to buy or sell shares.  
\subsection{Optimal Trade Execution}
Optimal trade execution refers to the process of buying or selling a financial asset while achieving the desired objective. Generally, this problem formulates as follows: Within a timeframe of $T$ timesteps, ${0,1,\cdots, T-1}$, a trader who possesses an inventory of $O$ shares must buy or sell the entirety of the inventory. 
For each timestep $t\in \{0,1,\cdots, T-1\}$, the trader determines the number of shares to order $O_t$ based on the information from LOB, and carries out the trade at an execution price $p_{t}$. 
For the group of traders seeking to maximize trading profit, their objective is to find a strategy that maximizes or minimizes the average execution price $\bar{P}$, which is calculated as 
$$\bar{P} = \sum^{T-1}_{t=0}\frac{O_t}{O}p_t.$$
While many financial firms pursue profits through trading, there are also numerous firms within the financial industry that prioritize consistently tracking the VWAP over maximizing trading profit. The focus of our paper is tailored for  this second group of traders. The VWAP is calculated as 
$$\text{VWAP} = \frac{\sum^{T-1}_{i=0}p_tq_t}{\sum^{T-1}_{i=0}q_t},$$
where $q_t$ is the execution volume determined by the market. Therefore, the primary objective of this paper is to minimize the difference between the VWAP and $\bar{P}$, using a dual-level approach.

\section{Related Works}
\subsection{Optimal Execution without RL}
In \cite{bertismasandlo1998}, a fundamental work for optimal trade execution is proposed, where the authors assume that market prices follow an arithmetic random walk.
They use a dynamic programming principle to find an explicit closed-form solution. Building on this work, \cite{hubermanandstanzl2005} and \cite{almgrenandchriss2001} extended the result in \cite{bertismasandlo1998} by incorporating transaction costs, more complex price impact functions, and risk aversion parameters, under the assumption that market prices follow a Brownian motion.
These dynamical approaches, however, are difficult to apply directly in the real world due to the discrepancy between their strong market assumptions and the complexity of actual market conditions.

On the other hand, both practitioners and researchers have widely used the time-weighted average price (TWAP) strategy and the volume-weighted average price (VWAP) strategy (\cite{berkowitzetal1988,kakadeetal2004}), which are based on either pure rules or statistical rules.
Especially in \cite{kakadeetal2004}, the authors used historical data to estimate the average volume traded for each time interval and split the order accordingly. However, this strategy is not well-suited for capturing unexpected volatility.
\subsection{Optimal Execution with RL}
From the perspective that optimal execution problems are a type of sequential decision-making task, Reinforcement Learning (RL) has been commonly applied in this field. To the best of our knowledge, \cite{nevmyvakaetal2006} is the first work to leverage RL frameworks such as Q-learning \cite{watkinsanddayan1992} to optimal trade execution problems. It is important to note that the curse of dimensionality in Q-learning makes it challenging to handle high-dimensional data. 
While \cite{hendricksandwilcos2014} attempts to combine RL with the Almgren-Chriss model, it is a challenging task as the model depends on certain assumptions about the market dynamics.

Thanks to the advancements in deep RL, recent studies such as \cite{ningetal2021} and \cite{linandbeling2019} have utilized Deep Q-Networks (DQNs) \cite{mnihetal2013} for optimal trade execution, addressing the challenges of high-dimensional data and the complexity of the financial market without relying on any market assumptions. 
However, these methods require the design of specific attributes, which can be labor-intensive. 
Recently, several studies have investigated the use of proximal policy optimization (PPO) \cite{schulmanetal2017} based optimal execution frameworks, which do not require manually designed feature engineering. For instance, \cite{linandbeling2020}, \cite{fangetal2021} and \cite{panetal2022} have explored this approach in different scenarios. Specifically, \cite{panetal2022} focused on optimal execution with limit orders, while \cite{linandbeling2020} and \cite{fangetal2021} used market orders.

Prior literature often considers a relatively short trading horizon or the entire day for RL implementation. However, VWAP approximation becomes challenging when a short time frame is used since the variations of financial data are often insignificant within short time intervals. Moreover, the naive approach of considering the entire day can put a strain on the network architecture due to the long sequence length. To address these issues, we propose a dual-level approach method to enhance the accuracy of approximating the daily cumulative VWAP. In our approach, we make use of the fact that practitioners usually utilize the statistical U-shape property to divide the total order for the day in order to track the daily cumulative VWAP.

\section{Method Description}
In this section, we describe our dual-level approach and formulation of the optimal trade execution problem under the lens of reinforcement learning. We also provide details on the architecture of the Transformer and LSTM neural networks and how they are integrated into our proposed method.

\begin{figure*}[!htbp]
\centering
 \includegraphics[width=14cm]{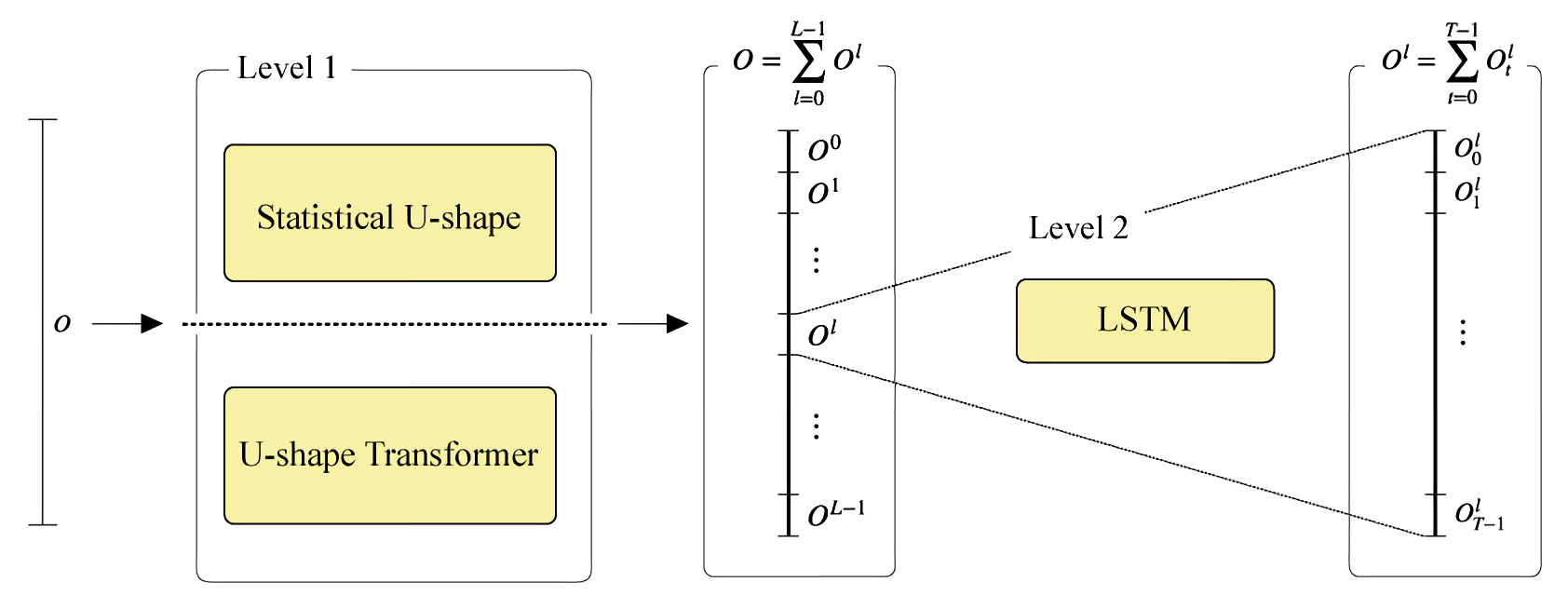}
   \hfil
\caption{A schematic illustration of our dual-level approach.}\label{architecture}
\end{figure*}
\subsection{Dual-level Approach}
Our approach involves two stages of distributing orders. In the first stage, we allocate the total daily orders $O$ into $L$ intervals. In the second stage, we further divide the orders from each interval into smaller executable orders using RL. 
Our approach begins by dividing the market hours of a day into $L$ intervals. For each interval $l$, we let $O^l$ as the total order to be executed, ensuring that the sum of all interval orders equals the total daily order, or $\sum^{L-1}_{l=0}O^l = O$. We suggest two methods for implementing the first stage. The naive approach is to adhere to the statistical U-shape trade volume distribution for each interval obtained from past data (e.g. the historical average of the past year). A more advanced method would be to use the output (predicted U-shape distribution) generated by a U-shape Transformer model.
In the latter case, $O^l$ is calculated progressively by referring only to market information from past days and previous intervals within the day.
For such an $l^{th}$ interval, which is comprised of $T$ timesteps, the LSTM model outputs $O^l_t$, which is the number of orders to be executed at timestep $t$ such that $\sum^{T-1}_{t=0}O^l_t = O^l$. 
Once $O^l_t$ is decided, it is averaged over $I$ execution steps in the corresponding $t^{th}$ subinterval. 
In other words, for each execution step, which is set as every 5 seconds, the final number of orders executed is $O^l_t/I$. 
The choice of averaging is supported by our findings that there is no significant difference in approximating the VWAP, regardless of how orders are distributed within the subintervals, as detailed in Section \ref{experiment_section}. 
A schematic illustration of this design is provided in Figure \ref{architecture}.

\subsection{MDP Formulation for Optimal Execution} \label{MDP_formulation}

In this section, we explain our Markov Decision Process (MDP) formulation of optimal trade execution. 
A MDP is typically represented using the tuple ($\mathcal{S}$, $\mathcal{A}$, $\mathcal{P}$, $r$, $\gamma$), where $\mathcal{S}$ is the state space, $\mathcal{A}$ is the action space, $\mathcal{P}$ is the transition probability, $r$ is the reward function, and $\gamma$ is the discount factor.
Our configurations of the state space $\mathcal{S}$ and action space $\mathcal{A}$ are similar to that of \cite{linandbeling2020}. 
It's worth noting that the MDP implementation in this paper is applied on a per $l^{th}$ interval basis for a given day.

\subsubsection{State}
The state $s^l_t \in \mathcal{S}$ consists of both public and private information. The public state is made up of the top 5 bid and ask prices, along with their associated volumes, while the private state includes the elapsed time and the current remaining volume to be executed.
The elapsed time ranges from $0$ to $T-1$ and the current remaining volume at timestep $t$ is equal to $O^l -\sum^{t-1}_{j=0}O^l_j$.

\subsubsection{Action}
Our LSTM model outputs a policy $\pi(\cdot|s^l_t) \in \mathbb{R}^{21}$, which is a discrete categorical probability distribution. From this, an action $a^l_t \in \mathcal{A}:=\{0,0.1,0.2,\cdots, 2\}$ is sampled to determine the number of orders to execute, with $O^l_t = a^l_t\frac{O^l}{T}$. To ensure that the total number of orders executed over the entire period is equal to $O^l$, that is $\sum^{T-1}_{t=0}O^l_t = O^l$, we impose the following two restrictions.

\begin{itemize}
    \item If $\sum^{j}_{t=0}O^l_t > O^l$ for some $j \in \{0, 1, 2, \cdots , T-1\}$, then $O^l_j = O^l - \sum^{j-1}_{t=0}O^l_t$
    \item $O^l_{T-1}$ is equal to the remaining volume to be executed at the last timestep for each interval (i.e. $0 \leq l \leq L-1$)
\end{itemize}
These restrictions are in line with the fact that it is essential for brokers to promptly acquire or liquidate all shares requested by their customers.


\subsubsection{Reward}
The reward function is designed to encourage the agent (LSTM model) to generate actions that result in orders that are close to the target volume $O_t^{l^*}$, in order to track the VWAP. This is achieved by using the price calculated from the executed orders.
Here, $O_t^{l^*}$ denotes the desired VWAP order at timestep $t$ in the $l^{th}$ interval such that $O^l = \sum^{T-1}_{t=0} O^{l^*}_t$. 
The daily VWAP can be obtained by 
\begin{equation}
VWAP_{day} = \sum^{L-1}_{l=0}\sum^{T-1}_{t=0}\frac{O^{l^*}_{t}}{O}p^l_{t},
\end{equation} 
where $p^l_{t}$ is the associated average market traded price of the corresponding $t^{th}$ subinterval. 

Our reward function compares $O_t^{l^*}$ and $O^l_t$ directly for a given $a^l_t$ as follows:
\begin{equation}
    r(a^l_t) := 
    \left\{\begin{array}{ll}
        1 & M^l_t < 0.01 \\
        0 & 0.01 \leq M^l_t < 0.05 \\
        -1 & \text{otherwise}
    \end{array} \right.
\end{equation}
where $M^l_t := \frac{|O^l_t - O^{l^*}_{t}|}{O^{l^*}_{t}}$.

Once MDP is determined, the goal of policy-based RL is to find an optimal policy parameter $\theta^*$, i.e.
\begin{equation}
\theta^*=\arg\max\mathbb{E}\left[\sum^T_{t=0}\gamma^t r_t|a_t\sim\pi_\theta(\cdot|s_t), s_{t+1}\sim \mathcal{P}(\cdot|s_t,a_t)\right],
\end{equation}
where $r_t = r(a_t)$.
\subsection{Neural Network Architecture and Training} \label{NN_Architecture_section}
We now turn to the architectural details of the Transformer and LSTM models and elaborate on how they are used in our dual-level approach. 
\subsubsection{Level 1: Transformer for Global U-shape Approximation}\label{sec-3-3-1}

\begin{figure}[h]
\centering
\includegraphics[width=7.8cm]{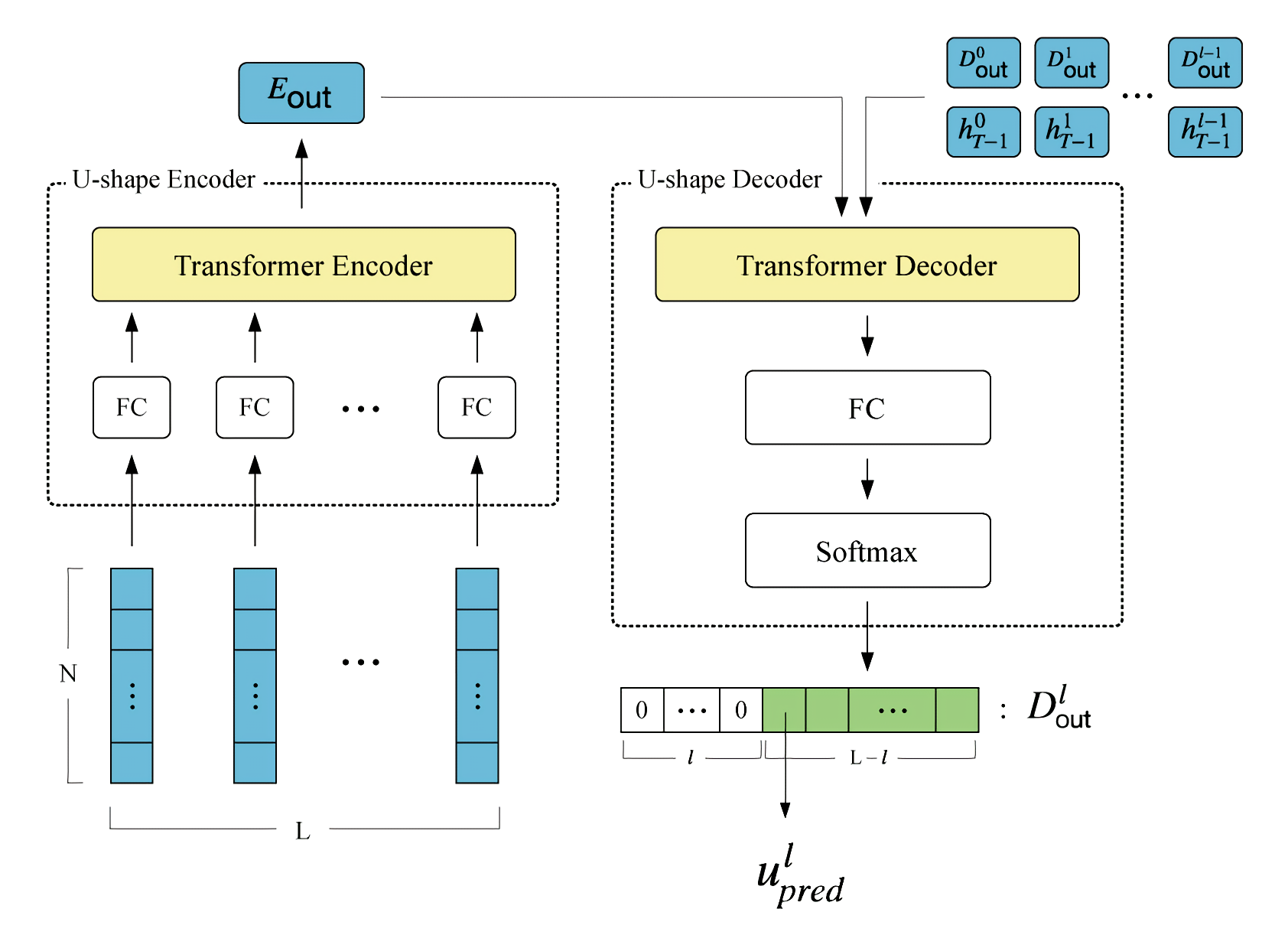}
   \hfil
\caption{The architecture of the U-shape Transformer. The U-shape Decoder in this figure represents the $l^{th}$ decoding step.} \label{Transformer}
\end{figure}

The objective of the Transformer model is to progressively predict the U-shape ratio for each interval in a given day. We use the structure of the Transformer Encoder and Decoder as proposed in \cite{vaswanietal2017}. The overall architecture of the Transformer model that we use is depicted in Figure \ref{Transformer}.

\textbf{\textit{U-shape Encoder.}} 
The U-shape Encoder $\mathcal{E}$ is composed of a Transformer Encoder and $L$ linear layers, all of which map to the same dimension. This module is responsible for learning the general dynamics of the U-shape distribution.
The input $E_{in}$ is a sequence of length $L$ of vectors each containing $N$ historical daily volume ratios of the corresponding interval in the sequence. The days from which these ratios are drawn are randomly selected. Each vector in the sequence is processed by a different linear layer that serves as an embedding to learn the unique features of each interval.
After passing through the embedding layer, the input sequence is then processed by the Transformer Encoder, resulting in the output $E_{out}$.

\textbf{\textit{U-shape Decoder.}} While $\mathcal{E}$ learns the general dynamics of the U-shape distribution, the U-shape Decoder $\mathcal{D}$ focuses on handling the day-to-day variations of the distribution. The U-shape Decoder is composed of a Transformer Decoder and $L$ linear layers, each mapping to different dimensions. In contrast to $\mathcal{E}$, these linear layers are applied to the output of the Transformer Decoder.

For every $l^{th}$ $(l > 0)$ decoding step, both $E_{out}$ and the cumulative input sequence are fed as input to the Transformer Decoder, $\mathcal{D}$. We denote the cumulative input sequence as $D^l_{in} = (D^l_{in,j})_{0\leq j\leq l-1}$, where $D^l_{in,j} := \text{CONCAT}(D^j_{out}, h^j_{T-1})$. Here, $D^j_{out}$ represents the output of $\mathcal{D}$ at the $j^{th}$ step, and $h^j_{T-1}$ is the last LSTM hidden vector from the $j^{th}$ interval.
The output from the Transformer Decoder at the $l^{th}$ decoding step is a vector of dimension $L + H$, which is then passed through a linear layer that reduces its dimension to $L - l$. The softmax function is applied to predict the ratios of the remaining $L - l$ intervals, and the first $l$ components are zero-padded.
Note that we use $h^j_{T-1}$ to capture the day-to-day variations in the U-shape distribution, as it holds interval-specific information for that particular day. To ensure $\sum^{L-1}_{l=0}u^l_{pred} = 1$, the final $l^{th}$ interval volume ratio prediction is calculated as 

\begin{equation}
    u^l_{pred} = \left(1 - \sum^{l-1}_{j=0}u^j_{pred}\right) \cdot D^l_{out}[l].
\end{equation}
It's worth noting that $D^0_{in}$ is constructed by applying an additional linear layer that transforms the dimension of the vector holding the top 5 bid/ask volume averages of pre-market data to $L + H$.

\begin{figure*}[!htbp]
\centering
\includegraphics[width=14cm]{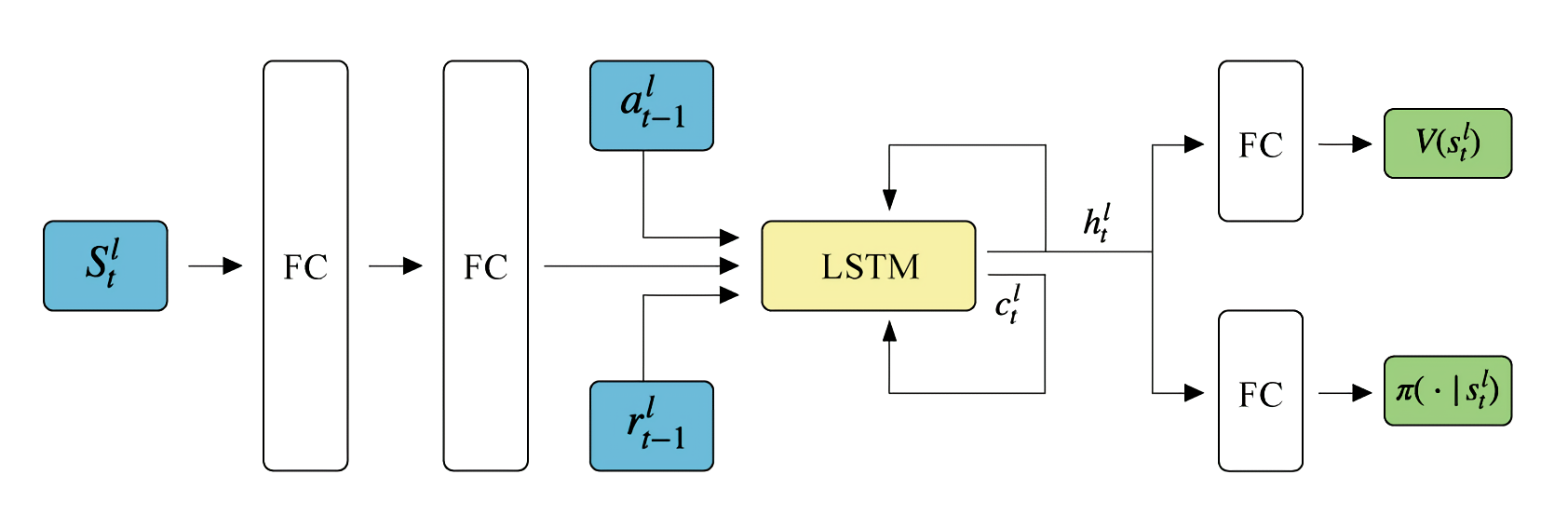}
\caption{The architecture of the LSTM model.} \label{LSTM}
\end{figure*}
\subsubsection{Level 2: LSTM for Local Order Distribution}\label{sec-3-3-2}
The objective of the LSTM model is to optimally allocate orders within the intervals of a given day. 
The structure of our LSTM model, which is outlined in Figure \ref{LSTM}, is similar to that of \cite{linandbeling2020}.
Given $s^l_{t}$ as input, it first passes through two linear layers with hidden dimension $128$ and is then concatenated with $a^l_{t-1}$ and $r^l_{t-1}:= r(a^l_{t-1})$. 
The resulting vector is then processed by an LSTM Cell with hidden dimension $H$ to obtain $h^l_{t}$, which is then passed separately through two linear layers to produce the policy $\pi(\cdot|s^l_{t})$ and the value $V(s^l_{t})$ respectively.
We note that the vector $h^l_{T-1}$ which is obtained at the last timestep is used as input to the Transformer Decoder. The action $a^l_{t}$, reward $r^l_{t}$, and the number of orders to execute $O^l_{t}$ is calculated as is described in Section \ref{MDP_formulation}. The LSTM Cell also receives $h^l_{t-1}$ and $c^l_{t-1}$ as input and its output includes $c^l_{t}$ as well.

\subsubsection{Training Procedure}
\textbf{\textit{U-shape Transformer}.} Since the goal of the first level of order allocation is to approximate the U-shape distribution while also tracking the daily cumulative VWAP, the training objective of the Transformer model is to minimize
\begin{equation}\label{e:loss01}
J_{TF}:=\mathbb{E}_{days}\left[\frac{c_1}{L}\sum^{L-1}_{l=0}\left(u^l_{true} - u^l_{pred}\right)^2 +c_2VAA\right]
\end{equation}
where
$c_1$ and $c_2$ are coefficients and 
$$\text{VAA} := \left|\frac{MP_{day} - VWAP_{day}}{VWAP_{day}}\right|$$
denotes the VWAP Approximation Accuracy (VAA), which we will further use as a general metric for performance evaluation in experiments, and 
$$MP_{day} := \sum^{L-1}_{l=0}\sum^{T-1}_{t=0}\frac{O^l_{t}}{O}p^l_{t}$$ is the price yielded by our model. 
This is one of the measures used in practice to gauge how close a trader has traded closer to their benchmark, the VWAP. 
If a trader has traded at a price that is exactly equal to the VWAP, then $VAA = 0$. 
Thus lower VAA would be better. 
The first term inside the expectation of \eqref{e:loss01} learns to make ratio predictions closer to ground-truth values. The second term penalizes the Transformer model if the final daily acquisition price generated by the combination of the Transformer and LSTM models deviates from the daily VWAP.
The primary objective of the U-shaped transformer in the first level is to distribute the daily total order with precision. However, relying solely on the first term of the loss function may not be enough to accomplish this objective. This is because $u^l_{true}$ does not consider the local information, such as the LOB, which can pose difficulties in tracking the daily VWAP accurately.
{Furthermore, if only the first term in \eqref{e:loss01} is taken into account, the reinforcement learning framework would not influence the first level, leading to independent operation of the first and second levels. To ensure that the micro information has an appropriate impact on the first level, we incorporate the second term into the loss function.}

\textbf{\textit{LSTM}}. To find the optimal parameter $\theta$ for the LSTM model, we use an actor-critic style PPO algorithm \cite{schulmanetal2017}, which is one of the most popular on-policy RL algorithms, as our base learner. One of the major advantages of the PPO in this task is its ability to adapt to changing market conditions and effectively learn from noisy and high-dimensional data. 
This algorithm uses the actor loss $J_{\text{PPO}}^{CLIP}(\theta)$ and the critic loss function $J_{\text{PPO}}^{VF}(\theta)$, which are defined as follows: 
\begin{equation}\label{eq-6}
\begin{split}
&J_{\text{PPO}}^{CLIP}(\theta) \\
&:=\mathbb{E}_t\left[\min\left(q_t(\theta)\hat{A}_t,\text{clip}(q_t(\theta),1-\varepsilon,1+\varepsilon)\hat{A}_t\right)\right],
\end{split}
\end{equation}
and
\begin{equation}
J_{\text{PPO}}^{VF} (\theta)= \mathbb{E}_t\left[(V^{\text{targ}}_t -V_\theta(s_t))^2\right], 
\end{equation}
where $q_t(\theta)= \frac{\pi_\theta(a_t|s_t)}{\pi_{\theta_{\text{old}}}(a_t|s_t)}$,
\begin{equation}
V^{\text{targ}}_t=\sum_{k=t}^{T-1} \gamma^{k-t}r_k \ \text{and} \ \hat{A}_t =V^{\text{targ}}_t -V_\theta(s_t).
\end{equation}
PPO's goal is to find a parameter $\theta$ that maximizes the main objective function $J_{\text{PPO}}(\theta) $, which is defined by
\begin{equation}
J_{\text{PPO}}(\theta) = J_{\text{PPO}}^{CLIP}(\theta) -c_3J_{\text{PPO}}^{VF} (\theta) + c_4\mathbb{E}_t[S[\pi_\theta](s_t)],
\end{equation}
where $c_3,c_4$ are coefficients, and $\mathbb{E}_t[S[\pi_\theta](s_t)]$ denotes an entropy loss term, where $S[\pi_\theta](s_t)$ is defined by
\begin{equation}
    S[\pi_\theta](s_t):=-\sum_{a\in\mathcal{A}}\pi_\theta(a|s_t)\log{\pi_\theta(a|s_t)}.
\end{equation}
By incorporating an entropy term, PPO can incentivize the agent to explore alternative actions, prevent getting trapped in suboptimal policies, and mitigate overfitting to the training data \citep{williams1992, mnihetal2016}.
To find the parameter $\theta$, PPO iteratively gathers episodes and updates the parameter $\theta$ using the following scheme:
\begin{equation}
\theta_{k+1} = \arg\max_{\theta}\mathbb{E}_{s}\mathbb{E}_{a\sim\pi_{\theta_k}(\cdot|s)}[J_{\text{PPO}}(\theta)],
\end{equation}
where $k$ stands for the $k^{\text{th}}$ step.
{The ``clip" operator in \eqref{eq-6} allows PPO to learn from previous experiences without becoming overly dependent on them. 
Moreover, PPO can effectively retain knowledge from past experiences while also acquiring insights from new experiences. 
This key attribute of PPO is one of its significant advantages.}
\section{Experiments}
Our dual-level approach significantly enhances the accuracy and consistency of tracking the daily cumulative VWAP, despite the challenges in improving the performance of RL frameworks for optimal trade execution in relatively short trading horizons. Furthermore, the U-shape Transformer proves efficient in capturing the daily variations of the U-shape distribution. In the following section, we conduct a series of experiments to verify these assertions.
\subsection{Experiment Settings}

\subsubsection{Implementation}
In all of our experiments, the hyperparameters for the loss functions, the length of the intervals, the number of total orders, and the trading horizon are fixed and set according to the following specifications:
\begin{enumerate}
\item[-]The coefficients for $J_{\text{TF}}$: $c_1 = 0.5, c_2 = 0.5$
\item[-]The coefficients for $J_{\text{PPO}}$: $c_3 = 1, c_4 = 0.01$
\item[-]The length of the intervals: $L = 19$, $T = 20$
\item[-]The number of total orders: $O$ $\sim$ $\mathcal{N}(2.5\times10^{-3}\mu, 6.25\times10^{-6}\sigma^2$), where $\mu$ and $\sigma$ is the average and standard deviation of the day total volume for the previous sixty days, respectively.
\item[-]Trading horizon of a day: $380$ minutes (From when the market opens at 09:00:00 to when it closes at 15:20:00).
\end{enumerate}

Since a typical day allows for 380 minutes of trading, our model first takes the daily total order, $O$, and breaks it down to $O^l$, 19 intervals of 20 minutes each. ($l \in [0,18]$) We either apply the statistical U-shape method or the Transformer model for this first stage allocation process. Then, within each interval, the LSTM model further subdivides $O^l$ into 20 sets of 1-minute subintervals, $O^l_t$. ($t \in [0, 19]$)

In each of these subintervals, $O^l_t$ is divided into equal portions and executed over a span of 12 steps, assuming that orders are executed every 5 seconds. The method and  hyperparameters used to train our agents for the experiment are summarized in Algorithms \ref{alg1}, \ref{alg2} and Table \ref{hyperparameter}, which are located at the end of the paper.

To conduct realistic simulations, we determine $O$ in a way that takes into account the volatility of daily volume, which differs from most previous works where it is fixed to a certain value. This choice aims to simulate the fluctuations of total orders that financial firms have to execute in a day, taking into consideration recent trade volume statistics for each stock. By considering the volume statistics for each stock, we account for the differences in total orders for various stocks, which is often encountered in real-world scenarios.
Additionally, during testing, we only use the daily ground truth U-shape volume ratios of dates in the training data for our Transformer Encoder to prevent looking ahead into the future.
As in previous works \cite{nevmyvakaetal2006}, \cite{hendricksandwilcos2014}, \cite{ningetal2021}, and \cite{linandbeling2020}, we make the following assumptions in our experiments.  
\begin{enumerate}
    \item We assume that the actions taken by our model only affect a temporary market, and that the market will recover to the equilibrium level at the next time step. 
    \item Commissions and exchange fees are ignored.
    \item Our model's orders are traded immediately without order arrival delays.
\end{enumerate}
We believe that the above assumptions are reasonable as we consider relatively small total orders in comparison to the daily total market volumes of the stocks.

\subsubsection{Datasets}
Our millisecond trade and limit order book (LOB) tick data is from the Korea Exchange (KRX). We use the daily trade and LOB data of Samsung Electronics (SE), SK Hynix (SH), Kia Corporation (KC) and POSCO Holdings Inc (PH) from January 1st, 2021 to December 31st, 2021. Our choice of the four firms in the KOSPI index is based on their liquidity and trade volume and variety in market capitalization and industry. In order to evaluate the effectiveness of our approach in relation to trading volume, we have opted to select stocks with a range of trade volumes (see Table \ref{volume}).
\begin{table}[h]
\centering
\begin{tabular}{ccccc}
\toprule
Stocks&SE& SH & KC  & PH  \\  
\midrule
Volume& 17,000K &3,800K&3,900K&430K\\
\bottomrule
\end{tabular}
\caption{Average volume for each stock.}\label{volume}
\end{table}
We divide the data into two sets, using January 1st to September 30th as training data and October 1st to December 31st as test data.

The millisecond raw data is preprocessed by first dividing them into groups of 5 seconds, and extracting data that represents each 5-second interval. 
We take the last LOB data of the 5-second interval to construct the MDP state and the 5-second VWAP and total traded volume for calculating the daily VWAP. 
The statistical U-shape statistics are constructed by taking the year average of the U-shape ratios of the 20-minute intervals in the day. The data utilized in generating $D^0_{in}$ shown in Section \ref{NN_Architecture_section} are the top 5 bid/ask volume averages of the raw pre-market data (from 08:30:00 to 09:00:00).
\subsubsection{Ablation Study}
We provide a brief explanation of all methods used for performance comparison during our experiments here. The hyperparameters for each method can be found in Table \ref{hyperparameter}.
\begin{table*}[h]
    \centering
    \begin{tabular}{ccccccc}
        \toprule
     Stocks&Metric&OPD&DQN& PPO  & HUL & TUL \\ \hline
        \midrule
      \multirow{3}{*}{SE}& Mean     & 26.27&11.53&12.00      &8.77  & \textbf{7.13}    \\
        &Standard Deviation  & 18.50&8.84&8.88          &  7.17 & \textbf{2.44}    \\
        &\% in 10bps& 25.00& 57.38&49.18&65.57&\textbf{88.52}     \\ \hline
       \multirow{3}{*}{SH}& Mean &38.55 &21.62     &     22.57       &\textbf{7.13} & 7.65\\
        &Standard Deviation  & 24.90&17.71& 16.92       &     \textbf{2.81} &3.63  \\
      &  \% in 10bps & 10.42 & 30.00&25.00        & \textbf{91.67} & 88.33     \\\hline
      \multirow{3}{*}{KA}& Mean     & 27.79&13.16&12.47              & 6.59 
 &  \textbf{6.03}   \\
        &Standard Deviation  &18.76&10.25& 11.55            & 8.36 & \textbf{3.30}     \\
      &  \% in 10bps & 18.75& 46.67&     56.67     &      \textbf{90.00} & 86.67  \\\hline
      \multirow{3}{*}{PH}& Mean & 32.44 & 14.84    & 14.64      & 22.52      & \textbf{12.27}   \\
        &Standard Deviation  & 20.89          & 13.47   & 12.63 & 47.11& \textbf{7.90}     \\
      &  \% in 10bps &   18.75   & 40.98 & 40.98   & 32.79 & \textbf{44.26} \\
        \bottomrule
    \end{tabular}
    \caption{Table \ref{tab:booktabs} summarizes the mean, standard deviation, and percentage of VAAs within the range [0, 10bps]. The Mean and Standard Deviation are represented in units of basis points (bps), and the $\%$ in 10 bps are represented in units of percentages.  For \textbf{OPD}, we excluded VAAs above the 80th percentile for fair comparisons since our training dataset was smaller than that used in \cite{fangetal2021}.}
    \label{tab:booktabs}
\end{table*}
\begin{enumerate}
    \item[-] \textbf{DQN} is the method proposed in \cite{ningetal2021}. To apply this model, \yj{we begin by dividing a given day into six 1-hour intervals, excluding the final 20 minutes of market hours to replicate the experimental setup in \cite{ningetal2021}. The total order for the day is then equally distributed amongst each hour, and \textbf{DQN} further allocates the order for five sub-intervals of 12 minutes each. The orders are executed equally for each 5-second time step within the 12-minute sub-intervals. To ensure fair comparisons, we evaluate the daily price generated by \textbf{DQN} against the daily cumulative VWAP calculated by excluding the final 20 minutes.}
    \item[-] \textbf{OPD} is the method proposed in \cite{fangetal2021}. \yj{To implement this model, we first divide a given day into ten intervals, each spanning 38 minutes, and then use \textbf{OPD} to allocate orders for each interval. Within each interval, the allocated order is equally distributed amongst each minute.}
    \item[-] \textbf{PPO} is the method proposed in \cite{linandbeling2020}. To apply this model, we equally distribute the day total order to each minute in the day, i.e. $O_j = O/380$ for all $j\in\{1,\cdots,380\}$ as in \cite{linandbeling2020}.
    \item[-] \textbf{(HU-)PPO} utilizes the statistical U-shape to determine $O^l$, and within each interval, the allocated order is equally divided among every minute, i.e. $O^l_t = O^l/20$ for all $t\in\{0,\cdots,19\}$, where PPO is finally used to distribute orders within the minute.
    \item[-] \textbf{HUL} is our proposed dual-level approach which uses the statistical U-shape to determine $O^l$.
    \item[-] \textbf{TUL} is our proposed dual-level approach that employs the Transformer model to determine $O^l$.
\end{enumerate}
\begin{figure}[ht]
\centering
 \includegraphics[width=7cm]{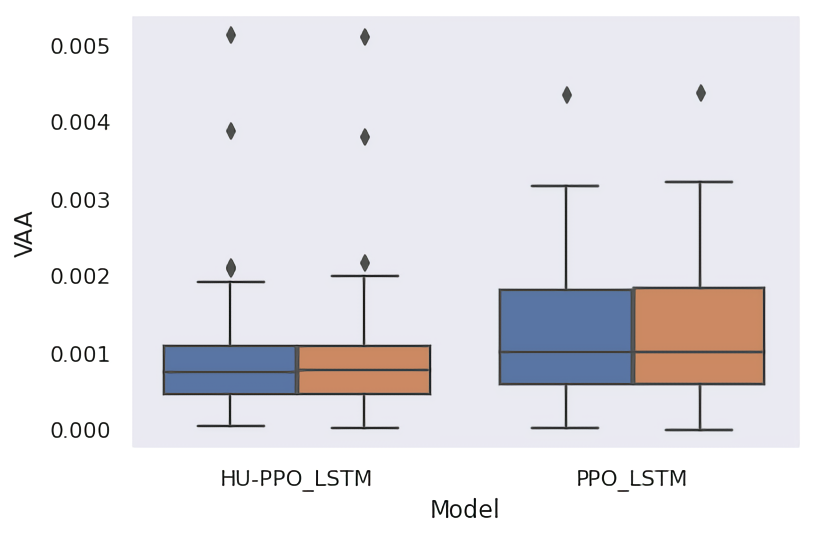}
   \hfil
\caption{VAA's of HU-PPO\_LSTM, PPO\_LSTM (in dark blue), and their alternative versions (in brown). In these alternative versions, orders are executed in the first step for every 1-minute trading horizon rather than using deep reinforcement learning.}\label{fig_worst_case}
\end{figure}
\begin{figure*}[htbp!]
\centering
\includegraphics[width=.75\columnwidth]{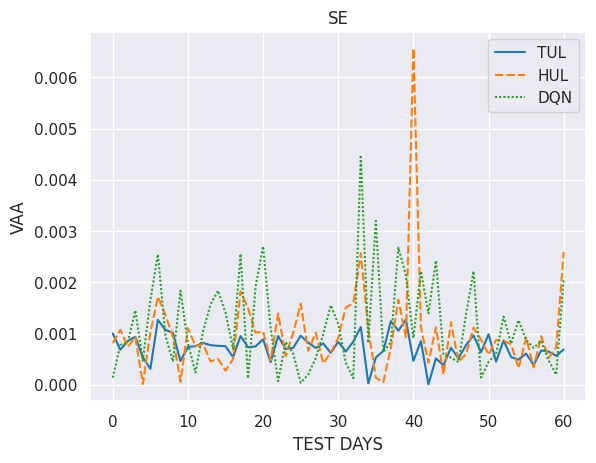} 
\includegraphics[width=.75\columnwidth]{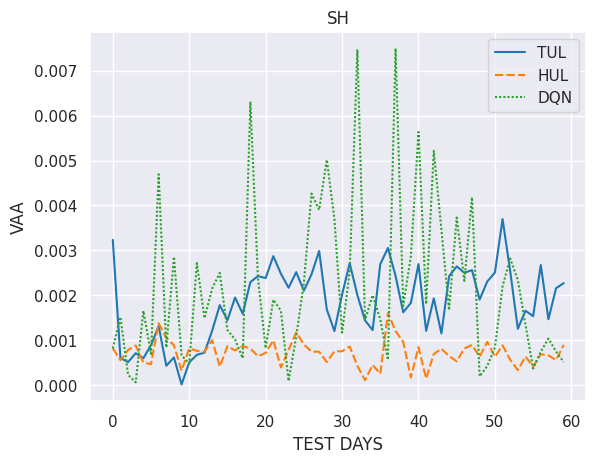}\\
\includegraphics[width=.75\columnwidth]{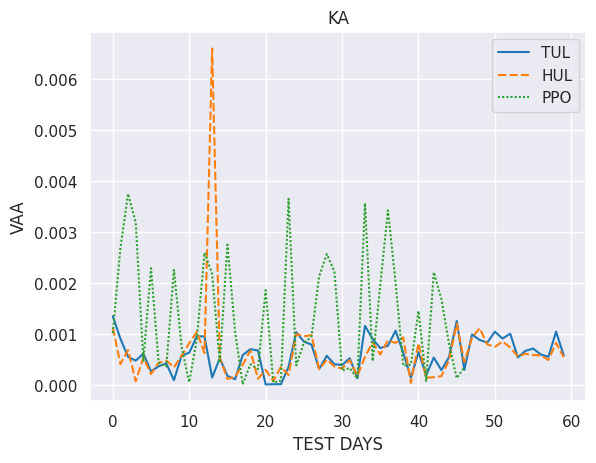}
\includegraphics[width=.75\columnwidth]{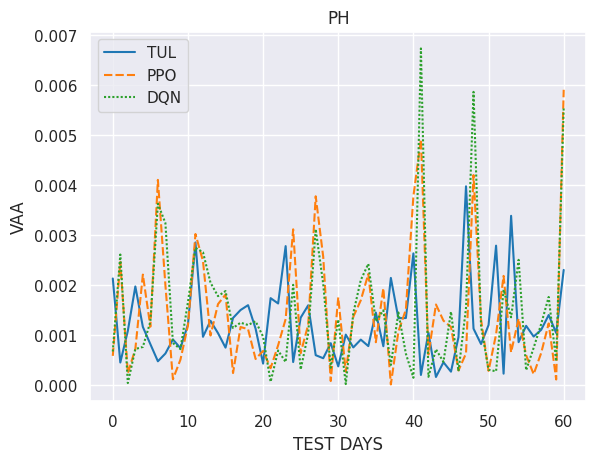}
\caption{{VAAs for the three best-performing models on the test data for each stock.}}\label{graph}
\end{figure*}
\begin{figure*}[htbp!]
\begin{tabular}{>{\centering\arraybackslash}m{0.3cm}>{\centering\arraybackslash}m{16cm}}
    \centering
    SE & 
    \includegraphics[width=15.5cm]{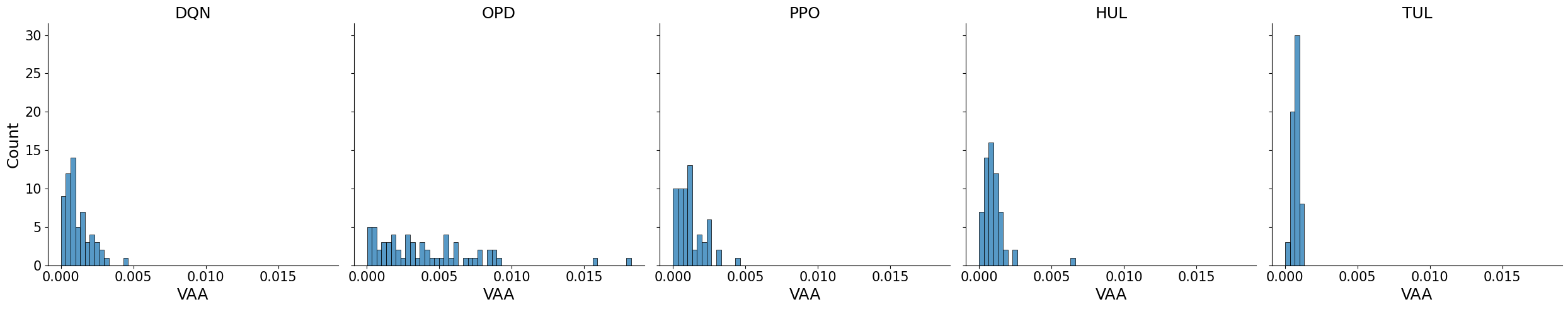} \\
    SH &
    \includegraphics[width=15cm]{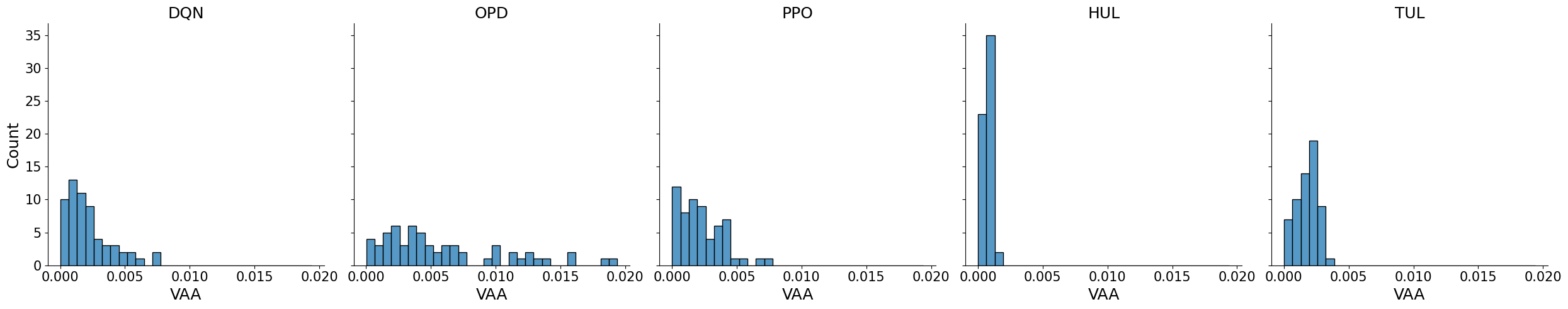} \\
    KC &
    \includegraphics[width=15cm] {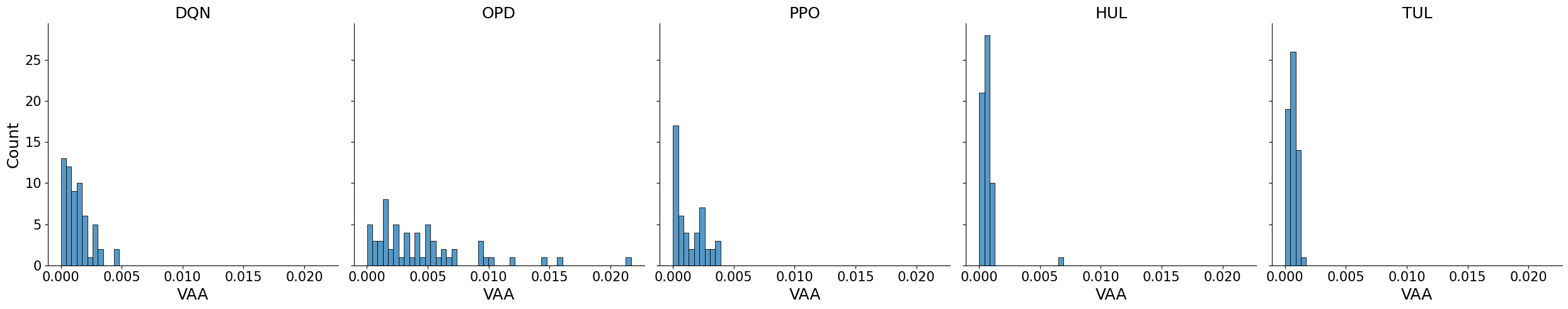} \\
    PH &
    \includegraphics[width=15cm] {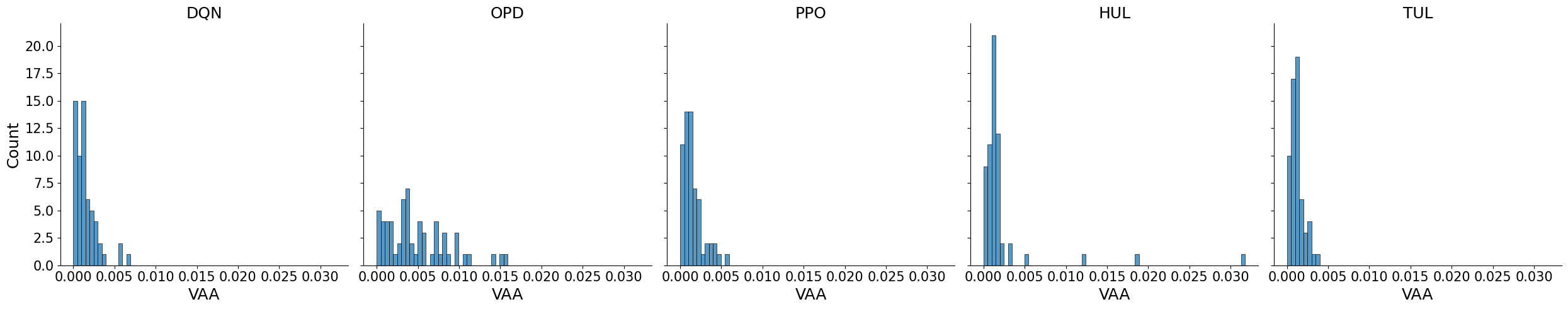} \\
\end{tabular}
\caption{The histograms of the VAA distributions for each model and stock on the test dates.}
\label{fig_samsung_histogram}
\end{figure*} 
\subsection{Experiment Results} \label{experiment_section}
We test our models, which are trained for 100,000 iterations, by conducting trading simulations in the buying direction. 
We first verify that using a short trading horizon for allocating orders degrades performance in approximating the daily cumulative VWAP. To do this, we investigate whether the decisions made by deep RL for the PPO and (HU-)PPO models actually contribute to lowering the VAA's produced during testing. Note that for both models, deep RL is using a 1-minute trading horizon. Thus, we intentionally construct a naive alternative version, in which the orders are simply executed in the first 5 seconds in the 1 minute trading horizon. The procedures taken to obtain the total order for the 1-minute interval for these versions are the same as their original counterparts. Figure \ref{fig_worst_case} displays the comparison of the VAA's on the test dates given by (HU-)PPO, PPO, and their alternative versions.

\begin{figure*}[htbp!]
\centering
\includegraphics[width=.7\columnwidth]{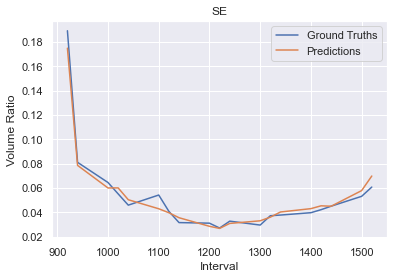} 
\includegraphics[width=.7\columnwidth]{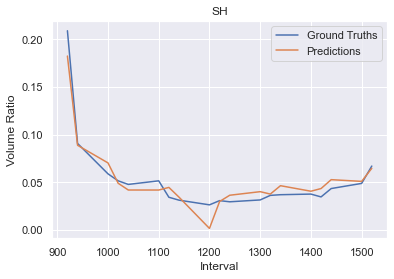}\\
\includegraphics[width=.7\columnwidth]{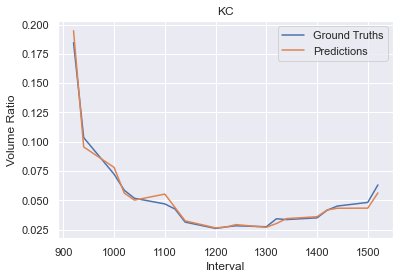}
\includegraphics[width=.7\columnwidth]{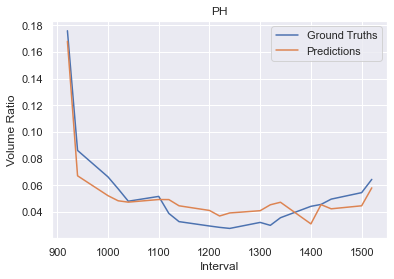}
\caption{{U-shape averages on test days for ground truths and Transformer-predicted values. }}\label{fig54}
\end{figure*}
\begin{figure*}[htbp!]
\centering
\includegraphics[width=.75\columnwidth]{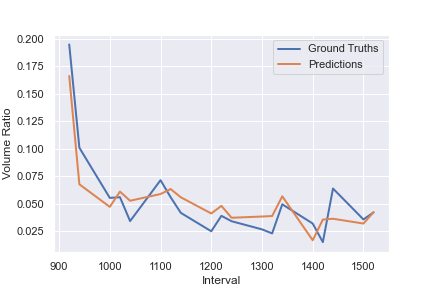}
\includegraphics[width=.75\columnwidth]{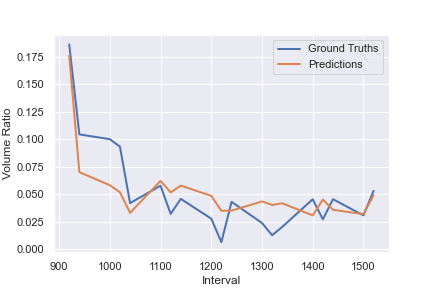}
\caption{{The ground truth and Transformer-predicted U-shape plots for two selected test dates for PH.}}\label{posco_u_shape_plots}
\end{figure*}

Both (HU-)PPO and PPO exhibit almost identical performance compared to their alternative versions, which suggests that in the context of VWAP the use of deep reinforcement learning does not significantly improve the approximation capability  if the trading horizon is too short. This implies that in order to effectively leverage the benefits of deep RL in sequential decision-making, it should be implemented on a longer trading horizon.


Secondly, we present Table \ref{tab:booktabs} and Figure \ref{graph} to demonstrate how our dual-level approach significantly improves performance. Table \ref{tab:booktabs} reports the mean and standard deviation values of VAA's produced by DQN, OPD, PPO, HUL and TUL across the test dates, and the percentage of them that fall under 10 bps. These methods are used by practitioners, to measure how the traders were performing relative to the VWAP. 
Our proposed dual-level approach demonstrates significant improvements compared to using short trading horizons and single-level neural networks. Although HUL occasionally yields better accuracy, TUL generally has much less deviation from the daily cumulative VWAP than all other methods. Especially for the stock PH, which is relatively less liquid compared to the other stocks, DQN and PPO show better mean and standard deviation values than HUL. However, TUL outperforms both models and demonstrates the best mean and standard deviation values. The histograms for VAA distributions given by DQN, OPD, PPO, HUL and TUL on each stock test data are drawn in Figure \ref{fig_samsung_histogram}, which depicts the clustering tendency of accuracies produced by TUL. 



Lastly, we demonstrate the effectiveness of the proposed U-shape Transformer using Figures \ref{fig54} and \ref{posco_u_shape_plots}. 
Figure \ref{fig54} shows the plots of the ground truth U-shape and Transformer-predicted U-shape values, where the U-shape ratio values for each 20-minute interval were averaged across all test dates. 
We observe that our U-shape Transformer is able to follow the U-shape pattern for the test dates on average for all stocks, which is in line with the comparable (and sometimes better) performance of TUL against HUL. 
This is further supported by the fact that HUL only utilizes the average U-shape ratio values of the past, while TUL approximates the U-shape distribution for the test dates accurately and uses predicted values for order allocation. 
Thus, TUL becomes more adaptive in tracking the daily cumulative VWAP as it incorporates predictions of fluctuations that may appear in the U-shape distribution in the future.

We also validate the performance improvements of TUL compared to other models, particularly HUL, by examining TUL's ability to capture daily fluctuations in the U-shape. 
Figure \ref{posco_u_shape_plots} displays the ground truth U-shape and Transformer-predicted distributions for two sampled test dates for the stock PH.
We deliberately choose PH since the performance gains of TUL when compared to HUL were the largest for this particular stock, which has daily volume distributions that deviates the most from the average due to relatively low liquidity. 
While the U-shape Transformer may occasionally make erroneous predictions, it quickly corrects itself and adapts to changes in the daily volume distribution, attempting to accurately follow ratio changes throughout the day.

\section{Conclusion}
In this study, we propose a dual-level approach to address the problem of tracking the VWAP with accuracy and consistency. Our model is unique in that it takes into account the well-known U-shaped intraday trading pattern. In the first stage, we allocate the number of orders to execute in each interval, taking the U-shape pattern into account. We consider two methods, statistical U-shape and U-shape Transformer, for implementing the first stage. The naive approach of using the statistical U-shape distribution improves performance. However, our experiments demonstrate that the U-shape Transformer performs even better. After the total daily orders are allocated to each interval in the first stage,  the LSTM model is used in the second stage to determine how the orders in each interval should be executed  Our simulation results show that the dual-level approach consistently and accurately tracks the daily cumulative VWAP.
\begin{table*}[htbp!]
\centering
    \begin{tabular}{p{4.8cm}| p{1.6cm}| p{1.6cm}| p{1.6cm} |p{1.6cm} |p{1.6cm}}
    \toprule
    \textbf{Hyperparameter} &DQN& OPD  & PPO& HUL &TUL \\
    \midrule
    \hline
    Outer iterations & 10000 & 10000 & 10000 & 10000 & 10000 \\
    \hline
    Inner iterations & 20 & 10 & 10 & 10 & 10 \\
    \hline
    Batch size & 50 & 10 & 20 & 10 & 10 \\
    \hline
    PPO CLIP $\varepsilon$ & - & 0.2 & 0.2 & 0.2 & 0.2 \\
    \hline
    Discount factor $\gamma$ & - & 1 & 1 & 1 & 1 \\
    \hline
    \makecell[l]{Number of trajectories \\ per each outer iteration}& 1000 & 160 & 200 & 152 & 228 \\
    \hline
    LSTM hidden dimension $H$& - & - & 128 & 128 & 129 \\
    \hline
    LSTM model learning rate&- &-  & 5e-5$\searrow$1e-5& 5e-5$\searrow$1e-5& 5e-5$\searrow$1e-5\\
    \hline
    DQN or OPD model learning rate & 1e-4 & 1e-4 & - & - & -\\
    \hline
    \makecell[l]{U-shape Encoder \\ input vector dimension $N$}&- & - & - & - & 20 \\
    \hline
    \makecell[l]{U-shape Encoder\\ embedding dimension} &- &- & - & - & 148 \\
    \hline
    \makecell[l]{Transformer Encoder \\ \& Decoder number of heads} &- &- & - & - & 4 \\
    \hline
    \makecell[l]{Transformer Encoder \\ \& Decoder number of layers} &- &- & - & - & 1 \\
    \hline
    \makecell[l]{Transformer Encoder \\ \& Decoder PFFN dimension}& -& - & - & - & 128 \\
    \hline
    Transformer model learning rate&- & - & - & - &1e-3$\searrow$2e-4 \\
    \bottomrule
    \end{tabular}
    \caption{Hyperparameters for DQN, OPD, PPO, HUL and TUL. {Note that $\searrow$ indicates a linearly annealing learning rate schedule. Network configurations of DQN and OPD follow \cite{ningetal2021} and \cite{fangetal2021}, respectively.}}\label{hyperparameter}
\end{table*}
\begin{algorithm*}
\caption{Dual-level Neural Network}\label{alg1}
\hspace*{\algorithmicindent} \textbf{Require}: Number of $outer$ $iterations$ and $num$ $days$
\begin{algorithmic}
    \STATE{Randomly initialize learnable parameters $\theta^\mathcal{E}$, $\theta^\mathcal{D}$, and $\theta^\mathcal{L}$}
    \STATE{Initialize trajectory buffer $\mathcal{T}$}
    \FOR{$j=1$ to $outer$ $iterations$}
        \FOR{$d=1$ to $num$ $days$}
            \STATE{Randomly choose a date in the training set}
            \STATE{$\mu\leftarrow$ the average of the daily volume for the last 60 days}
            \STATE{$\sigma \leftarrow$ the standard deviation of the daily volume for the last 60 days}
            \STATE{$\mathcal{T} \leftarrow \mathcal{T} \bigcup$ Gathering Episodes(date, $\mu$, $\sigma$)}
        \ENDFOR
        \STATE{Using $\mathcal{T}$, update $\mathcal{E}$, $\mathcal{D}$ by optimizing the loss function $J_{TF}$ in (5) with respect to $\theta^{\mathcal{E}}$ and $\theta^{\mathcal{D}}$}
        \STATE{Optimize the loss function $J_{\text{PPO}}$ in (9) with respect to $\theta^{\mathcal{L}}$}
        \STATE{$\mathcal{T} \leftarrow \emptyset$}
    \ENDFOR
\end{algorithmic}
\end{algorithm*}
\begin{algorithm*}
\caption{Gathering Episodes}\label{alg2}
\hspace*{\algorithmicindent} \textbf{Require}: The number of intervals in a day $L$\\
\hspace*{\algorithmicindent} \textbf{Input}: Date, $\mu$, $\sigma$ \\
\hspace*{\algorithmicindent} \textbf{Output}: Trajectory buffer of $i^{th}$ day $\mathcal{T}_{i}$ 
\begin{algorithmic}
\STATE{Initialize $\mathcal{T}_{i}$}
\STATE{$O\sim\mathcal{N}$($2.5$ $\times$ $10^{-3}\mu$, $6.25$ $\times$ $10^{-6}\sigma^2$)}
\STATE{$E_{out}=\mathcal{E}(E_{in})$}
\FOR{$l=0$ to $L-1$}
\IF{$l=0$}
\STATE{Construct $D^0_{in}$ from raw pre-market data}
\ELSE
\STATE{$D^l_{in} \leftarrow D^{l-1}_{in} \bigcup Concat(D^{l-1}_{out}, h^{l-1}_{T-1})$}
\ENDIF
\STATE{Obtain $u^l_{pred}$ and $D^l_{out}$ by procedure depicted in Section \ref{sec-3-3-1}}
\STATE{$O^l\leftarrow O\cdot u^l_{pred}$}
\STATE{Obtain $\tau_l=\{s^l_{t}, a^l_{t}, r(a^l_{t}), V(s^l_{t}), \pi(a^l_{t}|s^l_{t}), O^l_{t}\}_{t \in \{0, 1, 2, \cdots , T - 1\}}$ and $h^l_{T-1}$ by procedure depicted in Section \ref{sec-3-3-2}}
\STATE{$\mathcal{T}_{i} \leftarrow \mathcal{T}_{i} \bigcup \tau_{l}$}
\ENDFOR
\end{algorithmic}
\end{algorithm*}

\section*{Acknowledgments}
\noindent  
The work of Y. Hong was supported by Basic Science Research Program through the National Research Foundation of Korea (NRF) funded by the Ministry of Education (NRF-2021R1A2C1093579) and the Korea government (MSIT)(No. 2022R1A4A3033571).

\bibliography{reference}

\end{document}